**Number of words for the abstract: 323**

**Number of words for the main text: 5205**

**Number of words for the references: 2576**

**Number of words for the entire text: 8282**

# Title Page

**Desires and Motivation: The Computational Rule, the Underlying Neural Circuitry, and the Relevant Clinical Disorders**

Yu Liu[1,2], Yinghong Zhao[3] and Mo Chen[1,2]*

1. Jiangsu Province Key Laboratory of Anesthesiology, Xuzhou Medical University, Xuzhou, Jiangsu, China, 221004
2. Jiangsu Province Key Laboratory of Anesthesia and Analgesia Application Technology, Xuzhou Medical University, Xuzhou, Jiangsu, China, 221004
3. Department of Medical Imaging, Xuzhou Medical University, Xuzhou, Jiangsu, China, 221004

*Correspondence: Mo Chen, Email: chenmo@xzhmu.edu.cn

Present address: Jiangsu Province Key Laboratory of Anesthesiology, Technology Building D421, Xuzhou Medical University, 209 Tongshan Road, Xuzhou, Jiangsu, China, 221004

## *Short abstract*

Organism is a dissipative system. Therefore, dealing with deficiencies and keeping a balance are deemed as essential. If desires are defined as the absence of the ideal state, such an absence might be diverse. Due to the position difference of rewards and the spatial integrity of self-body, the organism can only pursue our desires in sequential manners. The exclusive motivation is formed correspondingly. Thus, the Desire-Motivation computation is very important for following actions. In this paper, a mathematical framework of Desire-Motivation computing was presented, which could dynamically regulate the motivation stability, motivation intensity and motivation decision time, and exhibited a close correlation with personality and psychiatry disorders.

## *Long abstract*

As organism is a dissipative system. The process from multi desires to exclusive motivation is of great importance among all sensory-action loops. In this paper we argued that a proper Desire-Motivation model should be a continuous dynamic mapping from the dynamic desire vector to the sparse motivation vector. Meanwhile, it should at least have specific stability and adjustability of motivation intensity. Besides, the neuroscience evidences suggest that the Desire-Motivation model should have dynamic information acquisition and should be a recurrent neural network. A five-equation model is built based on the above arguments, namely the Recurrent Gating Desire-Motivation (RGDM) model.

Additionally, a heuristic speculation based on the RGDM model about corresponding brain regions is carried out. It believes that the tonic and phasic firing of ventral tegmental area dopamine neurons should execute the respective and collective feedback functions of recurrent processing.

The analysis about the RGMD model shows the expectations about individual personality from three dimensions, namely stability, intensity, and motivation decision speed. These three dimensions can be combined and create eight different personalities, which is correspondent to Jung's personality structure theorem. Furthermore, the RGDM model can be used to predict three different brand-new types of depressive disorder with different phenotypes. Moreover, it can also explain several other psychiatry disorders from new perspectives.

**Key words:** Desire Motivation computation; Depressive disorder; Personality type; Mesolimbic dopamine system; recurrent neural network

# 1 Introduction

Using accurate physical and mathematical laws to explore and understand organism behaviors is a long-time goal set by neuroscientist, psychologist, behaviorist, and economist etc. Based on a commonly-known inference of Schrödinger that "life feeds on negative entropy."(Schrodinger 1944)(P32) to modern theories, e.g. free energy principle(Buckley, Kim, McGregor, & Seth 2017, Friston 2010), Bayesian brain theory(Doya, Ishii, Pouget, & Rao 2007, Knill & Pouget 2004), and reinforcement learning theory (Dayan & Daw 2008, Neftci & Averbeck 2019), scientists figured out a variety of frameworks to link the human behaviors to the computable dynamic processes regardless of how complex the process can be (Uttal 2020). In this paper, we started from the commentary of these theories about brain and then show some great power of the dynamical equations on linking neuroscience and part cognitive behaviors.

Organism is a dissipative system. Therefore, dealing with deficiencies (e.g. incremental entropy or low free energy) and keeping a balance are deemed as essential. Based on this point of view, it can be seen that information collection and identifying the enemy of balance are the first step of living organisms(Moller 2003).Factors leading to imbalance are called as needs or desires (Doyal & Gough 1984, Maslow 1958, Reiss 2002), which are used in this paper to represent the perceptual lack of material, energy, or ability to return to the best state or to approach the ideal state expected in the future. Although there are some theories claiming that desires and actions are directly linked together and there is a desire-action loop to keep the balance of body(Sinhababu 2017), there is a question that if we define desires as the absence of the ideal state, the absence may be diverse. Regarding the implementation of the human body, due to the position difference of reward targets and the spatial integrity of self-body, we usually can only pursue our desires in a sequential manner. Although debates regarding parallel versus serial desires do exist, for choices between water and food when they are in different rooms, desires do have to be satisfied in an order rather than simultaneously (Fischer & Plessow 2015, Townsend 1990)?

The terminology "motivation" was chosen in this work due to its widely application as the driving force of actions. It is also used for a more precise description of the choice of desires (Heckhausen & Heckhausen 2008, Nuttin, Lorion, & Dumas 1984). Together with action-plan and execution, we announced a precise step of human perception-action loop. Sensations will tell us our desires in a specific time. The brain calculates the exclusive motivation by using this desire information. An action plan can be figured out by combining the environment information and memory. Then people will finally execute the plan and get the prediction-reality error (positive or negative) feedback (Figure 1). The following part (2-4sections) shows some interesting achievements of the related dynamical equations under this framework, even though it is just a start and only focuses on the characteristics of the process from desires to motivation. Understanding the perception-action loop can move abundant theoretical and experimental brain studies forward. Under the computational theories of reinforcement learning framework, people choose their actions based on their value functions, which can also show how much future reward is expected from each action(Sutton & Barto 2018). The free-energy principle claimed that any self-organizing system that is at equilibrium with its environment must minimize its free energy(Friston 2010). All of the theories and frameworks above can help us to explore the details of the perception and action.

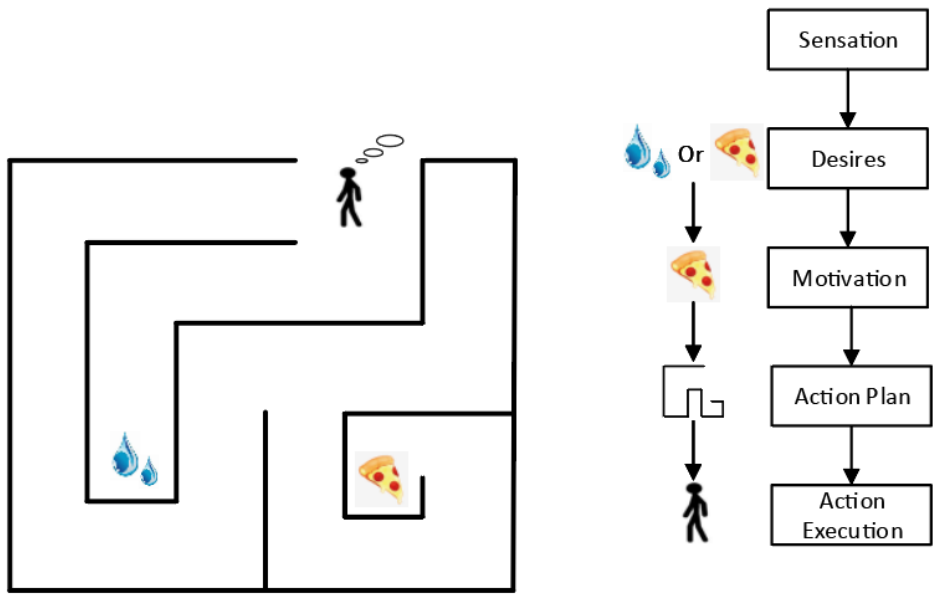

**Figure 1.    Demonstration of steps from sensation to action**

Dealing with deficiencies and keeping a balance are essential for a dissipative system. From sensation to action, there are five major steps involved, including sensations to desires, desires to motivation, motivation to action-plan, action-plan to action-execution, and the prediction-reality error feedback.

Our calculation is based on the desires of food, water, pain as well as some other desires. Human desires are quite extensive, including some noble purposes and goals. For example, belief and affection have more complex effects on human behaviors, which are not discussed in this paper. Besides, another computational process between desires and actions beyond motivation, such as the action-plan, is also not included in this article either.

## 2   The Performance Characteristics of Desire-Motivation Models

Although establishing rational and testable Desire-Motivation Models (DMM) is imperative, defining clear characteristics of DMMs is important as well.

## 2.1 Input and output of the model

No matter what form the Desire-motivation model has, it should have some basic characteristics. The first would be that the input of the model needs to involve a variety of human desires. These desires should have different intensity values, which are represented by the vector $\vec{D}$ in this paper.

As the output of the model, motivation should have the characteristics of Sparse Vector. Most values are 0 while only the values that need to be satisfied have their intensity values, which are represented by vector $\vec{M}$ in this paper. A question comes up here, which is how sparse $\vec{M}$ is. Although sometimes one action could lead to multiple rewards and satisfy multiple desires, one-by-one model is also acceptable.

Another aspect needs to be noticed is that the mapping from $\vec{D}$ to $\vec{M}$ is a continuous process of change. When we complete this mapping to a stable motivation and interact with the environment through behavior, the forms of desires will change correspondingly. The motivation will change as well. In conclusion, the Desire-Motivation model should have dynamic $\vec{D} - \vec{M}$ mapping.

## 2.2 stability of the model output

As the unit of Desire-Motivation computation, when and how the DMM system switches motivation is another important issue. The behavior of simple abstract organism is composed by the following major steps, including choosing the motivation, seeking action, taking action, desire change and re-choosing the motivation. If the motivation switches too frequently with the change of desires, then too much time will be spent on seeking action, which is not conducive to overall benefits. However, if a person always has stable motivation, focuses on pursuing one desire while ignores others, it is also appropriate. In the research, the stability of motivation is defined as attention(Pashler 1999). It is usually connected with school achievement and learning ability(Polderman, Boomsma, Bartels, Verhulst, & Huizink 2010, Schweizer & Moosbrugger 2004). In conclusion, the Desire-Motivation models should have appropriate stability instead of just being an unsophisticated real time winner-take-all model.

2.3 motivational intensity of the model output

Another aspect of the Desire-Motivation model to be considered is output (motivation) intensity, which has two practical implications. One is the individual aspect. Different demand intensities lead to different motivation intensities in the model. This basic assumption is of great significance for the subsequent behavior. Taking the desire for water as an example, when people are thirsty, people might choose the water resources close to them even though the water might not be that clean. However, if people are not thirsty, they might choose the water resources that are farther away. The balance of variability and stability has been well illustrated and studied. It is called as Yerkes Dodson law, which is a U-shaped curve of concentration and learning efficiency(Teigen 1994, Yerkes & Dodson 1908).The other is the population. The model needs controllable parameters. The same desires can produce different motivation intensities among different individuals. There is massive discussion about this in the prior studies regarding the motivation intensity. In conclusion, the Desire-Motivation models should have alterable output intensity, which vary from each other depending on people and situations.

In summary, it is argued in this paper that a proper Desire-motivation model should be a continuous dynamic mapping from the dynamic desire vector $\vec{D}$ to the sparse motivation vector $\vec{M}$. Besides, there should be some stability and adjustability of motivation intensity.

## 3  Structural Characteristics of Desire-Motivation models

The previous section demonstrates the schema characteristics of Desire-Motivation models. This section focuses on the structural characteristics that a Desire-Motivation model should have.

### 3.1  Active information acquisition vs passive information acquisition

The first question is how the external and internal information flows into the model. According to the neuroscience evidence, the information acquisition should be active rather than passive. Active information acquisition enjoys at least two benefits.

One benefit is the limited amount of information processed in a particular time. Like visual attention system, only the fovea has high resolution in retina. People use saccade and smooth pursuit to get active visual information(Hirsch & Curcio 1989, Niemeier, Crawford, & Tweed 2003). The visual information processing needs to go through a variety of filters in the brain (Buschman & Miller 2007, Hopfinger, Buonocore, & Mangun 2000, Itti 2000). The other benefit is information selectivity. People observe the surrounding environment with high information entropy through their eyes instead of just absorbing the information by talk (Lee & Stella 2000, W. Wang et al. 2011).

Evidence also shows that this process in the other external information collection organ (sense organ) is also active (Kriegeskorte & Douglas 2018, Schäfer & Zimmermann 2006). Meanwhile, there is also evidence about the internal information. For example, nociceptive information is controlled by inhibitory neurons located in the dorsal horn of the spinal cord in the well-known gate control theory of nociception. The top-down disinhibition signals would open the "gate" and evoked pain perception (Melzack & Wall 1965, Sufka & Price 2002). There are also similar regulatory mechanisms in the appetite-related circuits(Jo, Chen, Chua, Talmage, & Role 2006), olfactory bulb(D. Friedman & Strowbridge 2003, Lagier, Carleton, & Lledo), vestibular nuclei (S. M. Highstein & G. R. Holstein 2006) etc.. It is also argued in this paper that the Desire-Motivation model should also be an active information acquisition system with specific information gate controls.

### 3.2 Recurrent neural network

Essentially, an incentive computing system could be sketched as a network executing the winner-take-all functions. Such a system is often described as a decision-making network. Whether it is recurrent or feed-forward? Similar to the decision system in other brain regions, the current neuroscience believes that more recurrent neural networks are used in the brain than feed-forward neural networks. (Coultrip, Granger, & Lynch 1992, X. J. Wang 2008). Therefore, we argued that the Desire-Motivation model should also be a recurrent neural network rather than a feed-forward network.

### 3.3 A simple form of the Desire-Motivation model

Based on the evidence above, the Desire-motivation model should have dynamic information acquisition and should also be a recurrent neural network. Some light-hearted guidelines are finally obtained for building the basic form of the model. It is named as the Recurrent Gating Desire-motivation (RGDM) model (Figure 2). First, the inhibitory signal $Loc\_I$ of the information gate is constructed for active information acquisition. Similarly, motivation ($\vec{M}$) is an integrated function of desires ($\vec{D}$) and inhibitory gating ($Loc\_I$). Sigmoid is used here to ensure that the output is in the interval of 0 to 1(equation 1).

$$\vec{M} = Sig_1(\vec{D} - 0.5 \cdot Loc\_I) \tag{1}$$

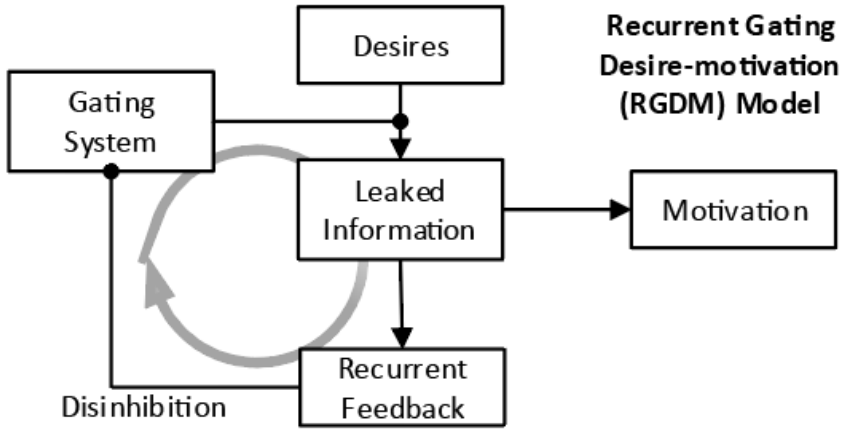

**Figure 2. Demonstration of RGDM model**

According to the RGDM model, active information and recurrent neural network are of great importance for a Desire-Motivation model. Local gating system controls the desire information leaking and a further network recurrently computes the most urgent desire and determines the motivation for following actions.

The output of the recurrent neural network, disinhibition signal ($Dis\_I$), is responsible for gate opening (equation 2).

$$Loc\_I = Sig(\varphi_2(0.5 - Dis\_I)) \qquad (2)$$

According to the classical recurrent neuron network, the control of $Dis\_I$ needs two signals. One is the respective recurrent feedback ($Res\_S$), which enhances (for selection) each processing cycle while the other one is the collective recurrent feedback signal ($Col\_S$), which controls the total amount of information(Camperi & Wang 1998, Sanzeni et al. 2020) (equation 3).

$$Dis\_I = \varphi_3 Res\_S + \varphi_4 Col\_S \tag{3}$$

Both $Res\_S$ and $Col\_S$ are the differential equations of computing motivation.

$$\frac{dRes\_S}{dt} = -\varphi_5 Res\_S + \varphi_6 \vec{M} \tag{4}$$

$$\frac{dCol\_S}{dt} = -\varphi_7 Col\_S + \varphi_8 (0.8 - \|\vec{M}\|) \tag{5}$$

Now, we have eight parameters ($\varphi_1$ to $\varphi_8$), which are further simplified into 4 (equation 6-10).

$$\vec{M} = Sig(10*(\vec{D} - 0.5 - Loc\_I)) \tag{6}$$

$$Loc\_I = Sig(10*(0.5 - Dis\_I)) \tag{7}$$

$$Dis\_I = \varphi_1 Res\_S + \varphi_2 Col\_S \tag{8}$$

$$\frac{dRes\_S}{dt} = \varphi_3(-Res\_S + \vec{M}) \tag{9}$$

$$\frac{dCol\_S}{dt} = \varphi_4(-Col\_S + (0.8 - \|\vec{M}\|)) \tag{10}$$

The subsequent sections show that 4 parameters have important biological significance. Besides, the basic performances of the model for the desire-motivation operation will also be analyzed (Figure 4-6).

## 4 RGDM model and neural circuitry

After determining the performance and structural characteristics of desire-motivation models, another problem that cannot be ignored is whether it is possible for the dynamic equation of RGMD model to support and explain the results of the neuroscience studies. The attempt is based on vast number of neurosciences studies, and inner link of the equations and the brain functions. The following part explains the relationship between dynamics and motivation behavior. In the RGMD model, four transfer variables are mainly used, including $Loc\_I$, $Dis\_I$, $Res\_S$, $Col\_S$. This section shows the corresponding relationship between these transfer variables and brain regions.

### 4.1 Where is the local inhibition ($Loc\_I$)?

For the Desire-Motivation model, the computation process in the brain needs sensory signals from the inside and outside of the body, such as the nociception for avoidance behavior(Julius & Basbaum 2001), the internal environment monitoring for feeding behavior(Sakurai et al. 1998), the vestibular receptor for balance (Wersäll, Flock, & Lundquist 1965), and the sexual perceptions (Argiolas & Melis 2003). Such receptors work around the clock and constantly monitor the deviation of the body from the ideal state. This vast amount of information is collected in the desire centers through the ascending pathways of the spinal cord, cranial nerves, and intracerebral pathways.

As discussed in section 3.1, our RGDM model introduces the active information acquisition mode, which requires vast local inhibitory gating system ($Loc\_I$). This raises a question, which is that "is there such a local inhibitory mechanism in the information acquisition system of the body?" The problem is complex and wide-ranging. However, it is undeniable that there are local inhibitory signals in the nuclei of multiple sensory systems, such as the dorsal horn of the spinal cord(Bardoni et al. 2013), the hypothalamus(Jennings, Rizzi, Stamatakis, Ung, & Stuber 2013), the vestibular nuclei(Stephen M Highstein & Gay R Holstein 2006). These local inhibitory neurons prove that the process of desire signal transmission to the center is not direct and continuous, but regulated.

### 4.2  Where is the brain circuit of desire-motivation computation?

There are abundant studies about the relationship between desires, rewards or motivation and the mesolimbic dopamine system, which is mainly referred to as the dopaminergic projection from the ventral tegmental area (VTA) to the nucleus accumbens (NAc) (Ikemoto , O'Connell & Hofmann 2011). This brain circuit is not only for physiological reward functions but also is closely connected to many major motivation related disorders, such as depression, attention deficit hyperactivity disorder (ADHD), anxiety, bipolar disorder and drug dependence (Berridge & Robinson , Cousins, Butts, & Young 2009,

Nestler & Carlezon , Radaelli et al. , Sturm et al. , Viggiano, Vallone, & Sadile), which will be further discussed in section 6.

Although people still have debates about the functions of the mesolimbic dopamine system, including "prediction error" theory, "incentive salience" theory and so on (Berridge 2012), almost all the theories are highly connected to the reward and motivation. According to the "prediction error" theory, the activity of VTA neurons miraculously agrees with the feedback signal in many reward-related tasks by coding "prediction errors" (Glimcher , Schultz). Meanwhile, the "incentive salience" theory mainly complained that the mesolimbic dopamine system showed a tight connection with the "incentive" control, and inactivation of this system would lead to deficits far more than learning dysfunction (Berridge 2007, Keiflin & Janak 2015). There are also studies proposing that the mesolimbic dopamine system should serve both functions (Saddoris, Cacciapaglia, Wightman, & Carelli 2015, Smith, Berridge, & Aldridge 2011) while the VTA tonic and phasic firings may play distinct, but essential roles (Bass et al. 2013, D. Chaudhury et al. 2013).

### 4.2.1 Where is the disinhibition center ($Dis\_I$)?

In our equation, $Dis\_I$ signaling is a descending disinhibition process. Expectedly, this signal may require several levels of transmission and regulations. However, the core issue is that $Dis\_I$ center activation will produce a variety of amplified desire signals. As the output center of mesolimbic dopamine system, does the NAc show this gating open function? It is indeed shown from the neuroscience studies that stimulation on NAc neuron will promote many desire-related behaviors, like feeding behavior(Narayanan, Guarnieri, & DiLeone 2010), sexual behavior(Balfour, Yu, & Coolen 2004), pain perception(Becerra & Borsook 2008) and even drinking behavior(Juarez et al. 2017).

### 4.2.2 Where is respective ($Res\_S$) and collective single ($Col\_S$) for Desire-Motivation computation?

It is mentioned in section 3.3 that there are two pathways that are essential for recurrent gating desire-motivation model, namely the respective ($Res\_S$) and collective ($Col\_S$) pathways. The input region of NAc, which is the ventral tegmental area (VTA), showed two firing patterns clearly, the tonic and phasic firing patterns (Morales & Margolis 2017). Some firing characteristics are also comparable to the RGMD model. In the RGMD model, the collective signal should exhibit a uniform response among all sensory modalities. According to recent studies, it is shown that the VTA phasic firings are homogenous among most individual neurons (Eshel, Tian, Bukwich, & Uchida 2016, Mohebi et al. 2019). Furthermore, to concurrently open or close all sensory gates, the firing pattern of the collective pathway should often be transient but not persistent. This conjecture tallies with the VTA phasic firing patterns

If a bold hypothesis is made, claiming that the major function of NAc-VTA circuitry is motivation control, multiple desire singles need to be taken while the single exclusive motivation that the body need to content at just this moment needs to be computed recurrently. Neuroscience-related research helps us to establish a complete framework of motivational and computational brain regions. However, the equations themselves should still be emphasized though they are still imperfect and crude in terms of the brain region correspondence. Nevertheless, if the imperfectness and crudity of the model does not conflict with the current large number of neuroscience, behavior and disease studies, the RGDM model is acceptable (Figure 3).

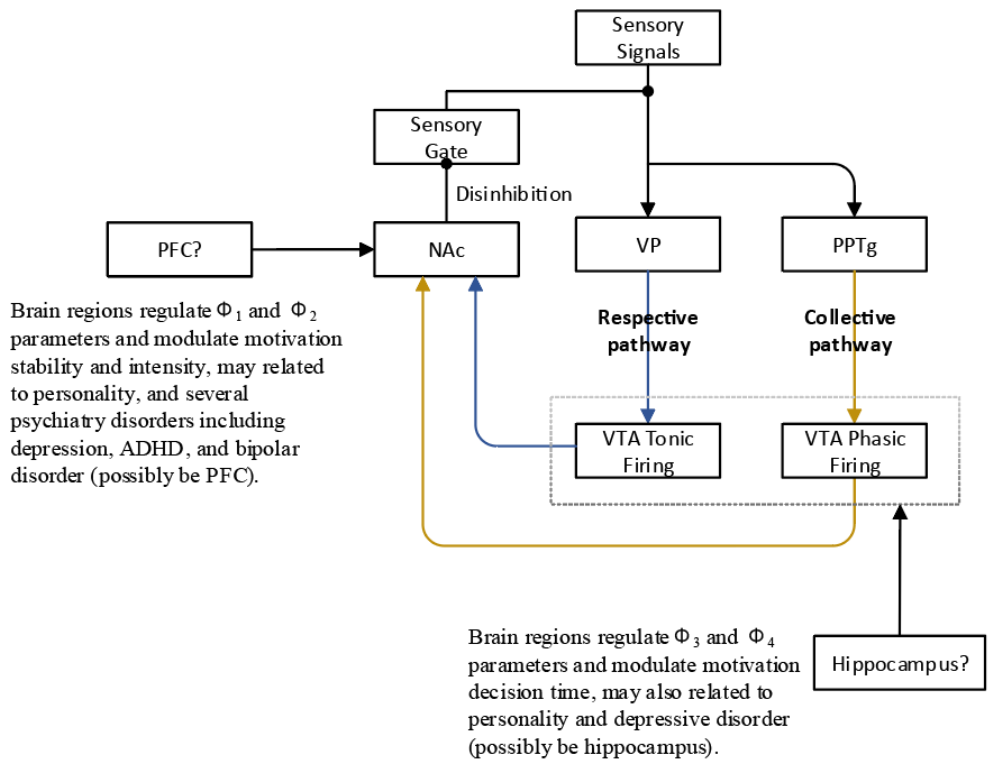

**Figure 3. RGDM model and possible brain regions**

The whole framework of the mesolimbic RGDM model depicts the involved brain regions and their interactions. Sensory signals from the sensation areas tell body the desires, but are gated by the local inhibition. VTA tonic and phasic pathways form respective recurrent and collective recurrent feedbacks. NAc is an integrated region and sends the disinhibition signal to the sensory gate. PFC and other related brain regions may regulate the motivation stability and intensity through controlling the VTA-NAc connection strength. Meanwhile, hippocampus and other related brain regions may regulate motivation decision time through controlling VTA tonic and phasic sensitivity. Abbreviations: PFC, the prefrontal cortex; PPTg, pedunculopontine tegmentum; NAc, nucleus accumbens; VP, ventral pallidum; VTA, ventral tegmental area.

# 5 RGDM Model: Motivation Control and Personality

As a network model that completes the continuous mapping from the dynamic desires vector $\vec{D}$ to the sparse motivation vector $\vec{M}$, different individuals need to have different input and output relationships. This part presents an analysis about the dynamic state of the RGDM model and also the parameters that change the properties of the whole network. Three aspects will be analyzed, including motivation stability, motivation intensity and motivation decision speed.

## 5.1 Parameter $\varphi_1$ and Motivation Stability

As mentioned in section 2.3, it is important to explore the stability of motivation first. After a proper cycle of iteration of the RGDM model, it eventually gives a relatively stable motivation. However, in reality, the input of RGDM model is constantly changing. Whether the chosen motivation needs to be maintained for a long time or not, and which parameters will affect the long-term stability of this motivation are important to individual differences.

In the RGDM model, four adjustable parameters were finally adopted. The simulation results showed that, the stability of motivation was dramatically influenced by the respective coefficient $\varphi_1$ (Figure 4) in the process of integrating respective and collective information. Based on the analysis in Section 4, we tend to correlate this parameter with the VTA tonic discharge intensity. Meanwhile, it can be regulated by the prefrontal cortex-NAc pathway. The correlation between the personality and the mesolimbic RGMD model is proposed in this part. Different individuals have different VTA tonic intensity or different NAc responses to VTA tonic discharge, resulting in different individual motivation stabilities.

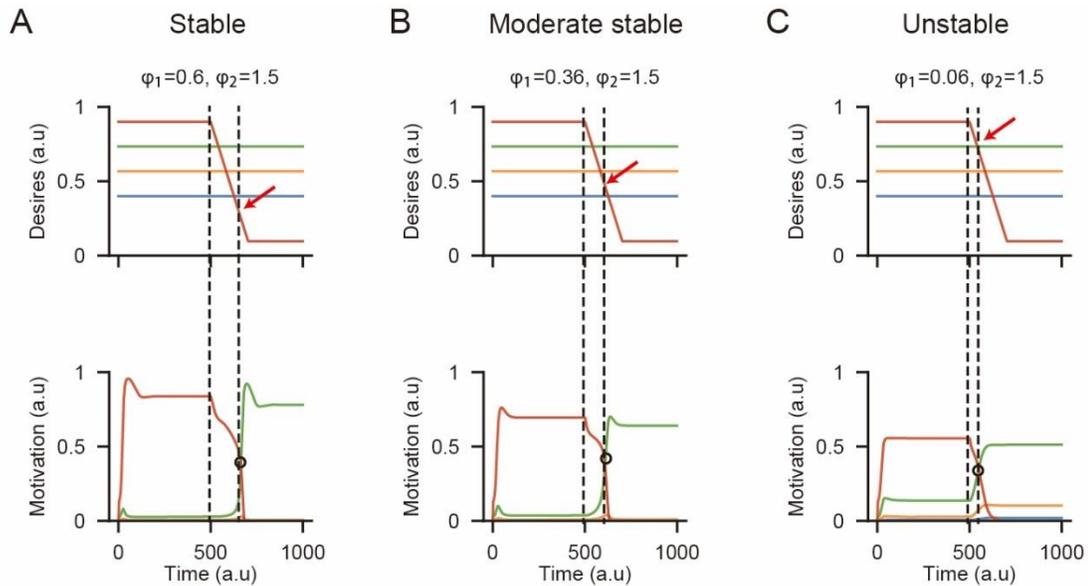

**Figure 4. RGDM model and motivation stability**

After incentive computation, the decreases of the winning desires (due to the intake of corresponding substance, red curves in the first rows of panels A, B and C) will elicit motivation transfers (denoted by open circles in the second rows of panels A, B and C). Under different parameter $\varphi_1$, the transfer will occur after different decreasing levels (denoted by red arrows in the first rows of panels A, B and C). (A) Stable state, the motivation transfer occurs only after large sensory input changes ($\varphi_1$=0.6). (B) Moderate stable state with $\varphi_1$=0.36. (C) Unstable state: the selected motivation is vulnerable to the changes of sensory inputs ($\varphi_1$=0.06).

### 5.2 Parameter $\varphi_1$ $\varphi_2$ and Motivation Intensity

Another important characteristic of the RGDM model is the output intensity of the motivation (under the same input premise). The stationary point of the RGDM model can be solved through the following equations:

$$M_{final} = (0.8\varphi_2 - Dis\_I_{final})/(\varphi_2 - \varphi_1) \tag{11}$$

$$M_{final} = Sig\left(10*\left(D - 0.5 - Sig\left(10*(0.5 - Dis\_I_{final})\right)\right)\right) \tag{12}$$

Through these two equations (equation 11 and 12), it can be inferred that for the same input desires, the output motivation intensity of the RGDM model is simply related to the parameters $\varphi_1$ and $\varphi_2$. Based on the analysis in section 4, the output intensity of the RGDM model is determined by the tonic discharge of VTA-NAc and the phasic discharge strength (Figure 5). Meanwhile, the motivation of strong desire and weak desire can be reduced synchronously by the regulation of the intensity of motivation and the reduction of tonic discharge. The reduction of phasic discharge can obviously reduce the motivation of strong desire than that of weak desire. The second assumption about personality classification based on the RGMD model is proposed. Different people will have different motivation strength levels corresponding to different VTA-NAc tonic or phasic firing strengths.

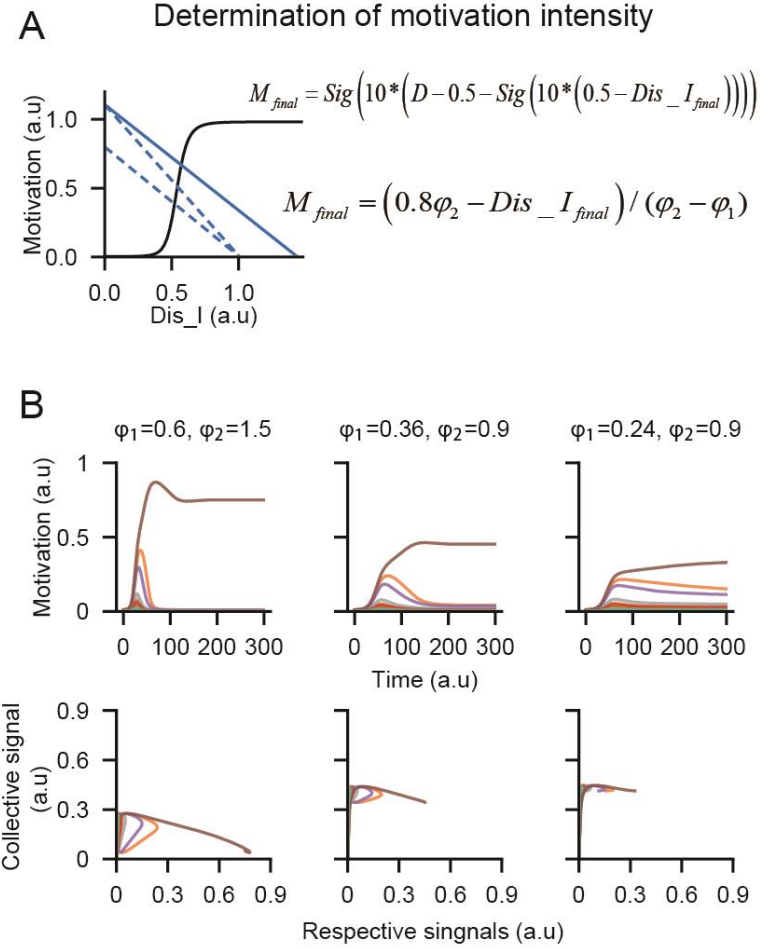

**Figure 5. RGDM model and motivation intensity**

(A) The stationary state of the RGDM model (the determination of the motivation intensity) is determined by the intersection point(s) of the blue lines (equations 7) and the black curve (equation 8). Low incentive state (the upper blue dashed line, corresponds to Figure 5B, middle column) will be caused by reducing $\varphi_2$ while keeping $\varphi_1/\varphi_2$ ratio constant. Any further decrease of the $\varphi_1/\varphi_2$ ratio will cause even lower incentive intensity (depression-like state) (the lower blue dashed line, corresponds to Figure 5B, right column). (B) At different parameter sets, the motivation intensity of the RGDM model differs (first row), and the evolving process (second row) also differs. From left to right, the incentive intensity decreases. Besides, the state represented by the lower right panel is depressive-like (very low incentive intensity).

## 5.3 Parameter $\varphi_3$ $\varphi_4$ and Motivation Decision Time

In the real life, the computation of both the inner and outer information is complicated. Our brain needs to collect ample information and sometimes need inference to make proper motivation decisions. The root(s) of the characteristic equation of equations 6-10 decide the decision time of RGDM model and the latency when the system accomplishes incentive computing. The oscillatory behavior of the RGDM model is decided by the following:

$$(1+\varphi_2\varphi_5 - (1-\varphi_1\varphi_5)(\varphi_3/\varphi_4))^2 - 4\varphi_1\varphi_2\varphi_5^2(\varphi_3/\varphi_4) \qquad (13)$$

The parameter $\varphi_5$ is the arbitrary parameter that is used when equations 6 and 7 are linearized. Interestingly, an independent influence factor $\varphi_3/\varphi_4$ on decision time of the RGDM model is observed, which represents the ratio of the sensitivity of VTA tonic pathway to the sensitivity of the VTA phasic pathways (Figure 6).

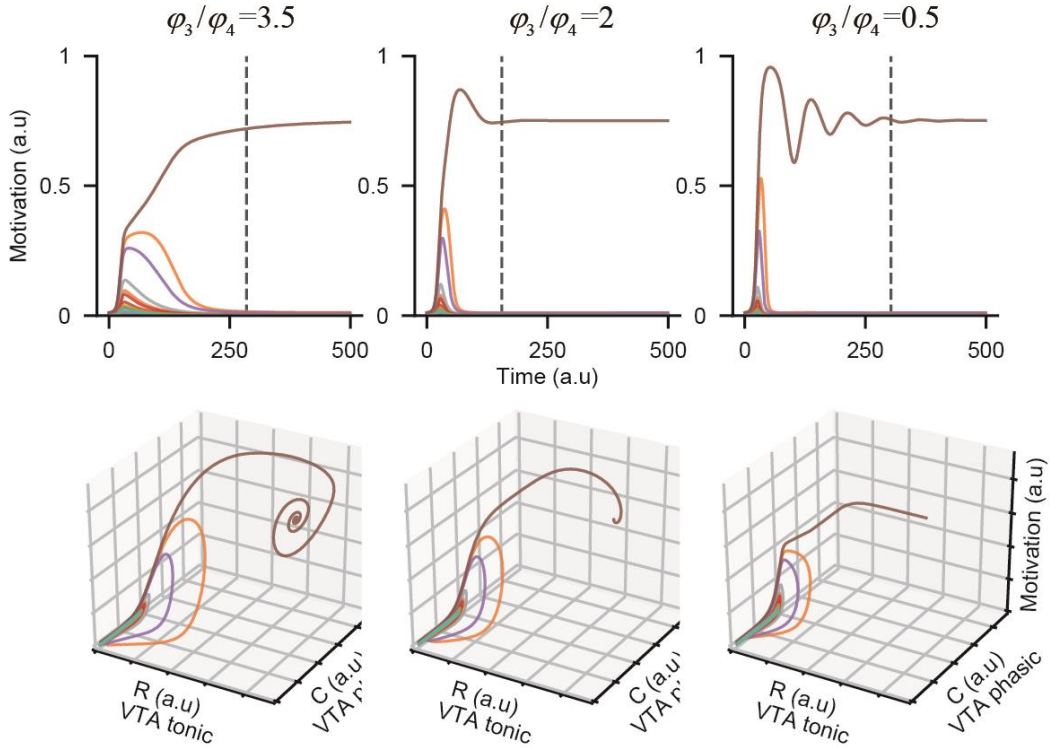

**Figure 6 RGDM Model and Motivation Decision Time**

Following the changes of the $\varphi_3/\varphi_4$ ratio (the ratio of the sensitivity of VTA tonic pathway to the sensitivity of the VTA phasic pathway), motivation decision time of the RGDM model varies from each other. When there is a large $\varphi_3/\varphi_4$ ratio, the model needs more time to make motivation decision (left column). When there is a proper $\varphi_3/\varphi_4$ ratio, the motivation decision is fast (middle column). When there is a small $\varphi_3/\varphi_4$ ratio, the motivation decision could be made fast, but needs time to become stable (right column).

For any chosen $\varphi_1$ and $\varphi_2$ parameter set, stable decision time in the RGDM model is a U-shape function of $\varphi_3/\varphi_4$. There is an inflection $\varphi_3/\varphi_4$, which results in the shortest decision time. However, we need to notice that when the $\varphi_3/\varphi_4$ ratio is small, the decision time is relatively long. However, when the $\varphi_3/\varphi_4$ ratio is too large, the system gives out unstable decisions instead of making no decision. This part presents an analysis regarding the third expectation of the RGMD model for personality classification, which claims that different people will have different motivation decision speeds corresponding to different VTA tonic/ phasic sensitivity ratios to the input signals.

In summary, the RGMD model shows more details about three dimensions of individual personality, namely stability, intensity, and motivation decision speed. These three dimensions can be combined and create eight different personalities. Interestingly, such a classification of personalities into eight categories has been proposed for a long time and already been used very widely(Geyer 1995, Jung 2016). Although further studies are needed, the correspondence of Jung's personality structure theorem and the recurrent gating desire-motivation model (Table 1) are used in this paper and can be used as the evidence.

**Table 1 Jung's Personality Classification and RGDM Model**

| Psychological type | Motive Stability | Motivation intensity | Motivation decision time | Possible parameter set |
|---|---|---|---|---|
| extroverted thinking | Steady | Strong | Fast | $\varphi_1 \uparrow \varphi_2 \uparrow \varphi_3/\varphi_4 \uparrow$ |
| extroverted feeling | Instability | Strong | Fast | $\varphi_1 \downarrow \varphi_2 \uparrow \varphi_3/\varphi_4 \uparrow$ |
| extroverted sensation | Steady | Weak | Fast | $\varphi_1 \uparrow \varphi_2 \downarrow \varphi_3/\varphi_4 \uparrow$ |
| extroverted intuition | Instability | Weak | Fast | $\varphi_1 \downarrow \varphi_2 \downarrow \varphi_3/\varphi_4 \uparrow$ |
| introverted thinking | Steady | Strong | Slow | $\varphi_1 \uparrow \varphi_2 \uparrow \varphi_3/\varphi_4 \downarrow$ |
| introverted feeling | Instability | Strong | Slow | $\varphi_1 \downarrow \varphi_2 \uparrow \varphi_3/\varphi_4 \downarrow$ |

| introverted sensation | Steady | Weak | Slow | $\varphi_1 \uparrow \varphi_2 \downarrow \varphi_3/\varphi_4 \downarrow$ |
| introverted intuition | Instability | Weak | Slow | $\varphi_1 \downarrow \varphi_2 \downarrow \varphi_3/\varphi_4 \downarrow$ |

# 6 RGDM Model: Motivation Control and Mental Diseases

When it comes to the basic characteristics of Desire Motivation model, the output of motivation should have the characteristics of stability, intensity, and decision speed. We realized that these basic characteristics are not only related with individual personalities, but also are implicative for the etiology of some psychoneurotic diseases. For example, mood disorders, attention deficit hyperactivity disorder, procrastination and other psychiatry disease are all identified to be related with the motivated choice.

## 6.1 Depressive disorder: Mood Disorder or Motivation Disorder?

Depressive disorder is a common issue all over the world, which is characterized by negative self-evaluation, low motivation, and somatic symptoms(Belmaker & Agam 2008, Uher, Payne, Pavlova, & Perlis 2014). Most of the time, people think of self-evaluation as the pathogeny for low motivation. However, it is important to notice if there are other possibilities. Low motivation also could be identified as the direct cause of depressive disorder. RGDM model at least can be used to explain the demotivation of depression. As it is shown in section 5.2, parameters could dramatically change the motivation intensities.

## 6.2 A new classification of depression based on the desire motivation model

The RGDM model can be used for the prediction of three independent factors, which will lead to extremely low motivation condition: the VTA tonic firing strength ($\varphi_1$), the VTA phasic firing strength ($\varphi_2$) and the VTA tonic/phasic sensitivity of inputs ($\varphi_3/\varphi_4$). There are three brand-new different types of depressive disorder with different phenotypes obtained (Table 2). There are various factors evoking these, including reuptake problem, secretion problem or other brain area control problem (Brebner et al. 2005, Laasonen-Balk et al. 1999, Pizzagalli et al. 2008, Raskin & Durst 2010). Interestingly, based on the RGDM model abnormally increase or decrease VTA phasic firing rate ($\varphi_2$ down or up regulation) could both lead to depressive-like behaviors, moreover both phenomena were observed (Dipesh Chaudhury et al. 2013, Ebner, Roitman, Potter, Rachlin, & Chartoff 2010, A. Friedman, Friedman, Dremencov, & Yadid 2008, Kumar et al. 2008).

**Table 2 Different phenotypes and neural mechanisms of three types of depression**

|  | unstable depression | stable depression | selective Depression |
|---|---|---|---|
| **pathogeny** | $\varphi_1 \downarrow$ or $\varphi_2 \uparrow$ | $\varphi_2 \downarrow$ | $\varphi_3/\varphi_4 \downarrow$ |
| **Motivation intensity** | Weak | Weak | Indeterminacy |
| **Motive Stability** | Instability | Indeterminacy | Indeterminacy |
| **Motivational choice time** | Indeterminacy | Indeterminacy | Long |

6.3  Other psychoneurotic diseases

Beyond depressive disorder, there are some other psychiatry disorders that can apply the RGDM model, such as the Attention Deficit Hyperactivity Disorder (ADHD), the major symptom of which is low motivation stability. Bipolar disorder is correlated with high motivation intensity while anxiety and procrastination are correlated with slow motivation decision speed. Evidence also supports that these disorders are highly correlated with the mesolimbic dopamine system (Alexander & Farrelly 2018, Arnsten 2009, Ashok et al. 2017, Avery, Clauss, & Blackford 2016, Calhoon & Tye 2015, Drevets 2000, Mehta, Monegro, Nene, Fayyaz, & Bollu 2019, Schlüter et al. 2019, Tovote, Fadok, & Lüthi 2015).

# 7 Some other interesting evidence and analysis:

RGDM model gives us an insight about the motivation related behaviors, and more questions are raised.

## 7.1 Motivation decision and Monte Calro Tree

It is very hard to make a proper decision both for the organism and artificial intelligence agent. Monte Calro Tree searching method is one of the methods that are proved to be effective (Hammersley 2013). Thinking deeper is deemed as helpful for competitions (Silver et al. 2016). this raises a question about where is the region of this recurrent deep thinking. This is still an open question. However, it must be related with the Desire-Motivation decision system. Is it hippocampus? We proposed it since some evidence showed that hippocampus not only shows where are we now but also can predict where are we in the future (Buckner 2010, Frank, Brown, & Wilson 2000, Lisman & Grace 2005).

### 7.2 Differences between men and women

Another features is that the dopamine release differs between men and women (Munro et al. 2006). According to RGDM model, this may represent some interesting features, motivational stability and motivation intensity whereas. Thus women might show a higher rate of depressive disorder (Piccinelli & Wilkinson 2000) while men are more easily influenced by drug addiction (Bobzean, DeNobrega, & Perrotti 2014). This difference in motivational stability and intensity may be having benefits on evolution and courtship behavior.

### 7.3 The influence of long-time goal

Another interesting phenomenon is that when we are doing simulation, a long-time goal is added to the system (week but long-lasting desire). Meanwhile, an autogenic regularity behavior pattern comes up. The system will deal with the long-time goal for a while and then deal with all the other desires intensively. Although it still needs further verification, a regularity life might not be depends on self-control but stem from a permanently life goal(Monk, Petrie, Hayes, & Kupfer 1994).

### 7.4 Conservation of evolution

If the motivational choice is mesolimbic, then the characteristics of the system, especially tonic and phasic firing, should be very conservative during the evolution process. We can notice the dopamine system in fish(Messias, Paula, Grutter, Bshary, & Soares 2016),

amphibians(Gonzalez & Smeets 1991), insects, (Van Swinderen & Andretic 2011)and even the earliest multicellular organisms(Van Swinderen & Andretic 2011)  (Chung & Spencer 1991a, 1991b)

## 8  Questions could be verified

RGDM model provides vast allowing quantitative testable inferences. These are several selected issues that need further verification.

a) The correlation between RGDM model parameters and motivational stability, motivational intensity, and motivation decision time.
b) The correlations between neuron response and behavior that RGDM Model predicted
c) Depression classification predicted by RGDM model.
d) Gender differences in the RGDM model.

## 9  Conclusion

In conclusion, we consider that an appropriate desire-motivation computation procedure is necessary for the organism. It is important to ensure that the Desire-Motivation kinetic model have some basic characteristics, such as proper stability, proper intensity as well as some other basic features. From the existing results of computational neuroscience, Desire-Motivation model may have both gating and recurrent characteristics, and can correspond to the existing neural pathways and research results of mesolimbic dopamine system. A simple RGDM model can provide three different personality dimensions, including motivational stability, motivational intensity, and motivational decision speed, which are consistent with the findings of the previous studies about personality. In terms of the disease, the RGDM model may provide three new subtypes of depressive disorder. However, it still needs further experimental verification for the conclusions.

**Competing financial interests:**

The authors declare no competing financial interests.

**Acknowledgements**

This study was supported by the following foundations: National Natural Science Foundation of China (Nos. 31400957 to Y.L., and 31700903 to M.C.), Natural Science Foundation of Jiangsu Province (Nos. BK20140218 to Y.L. and BK20150208 to M.C.), Talent Young Foundation of Xuzhou Medical University (Nos.D2014012 to Y.L. and D2014013 to M.C.) and Qing Lan Project to Y.L.

Argiolas, A. & Melis, M. (2003) The neurophysiology of the sexual cycle. *Journal of endocrinological investigation 26*(3 Suppl): 20.

Arnsten, A. F. T. (2009) The Emerging Neurobiology of Attention Deficit Hyperactivity Disorder: The Key Role of the Prefrontal Association Cortex. *J Pediatr 154*(5): I-S43.

Ashok, A. H., Marques, T. R., Jauhar, S., Nour, M. M., Goodwin, G., Young, A. H. & Howes, O. D. (2017) The dopamine hypothesis of bipolar affective disorder: the state of the art and implications for treatment. *Molecular psychiatry 22*(5): 666-679.

Avery, S., Clauss, J. & Blackford, J. (2016) The human BNST: functional role in anxiety and addiction. *Neuropsychopharmacology 41*(1): 126.

Balfour, M. E., Yu, L. & Coolen, L. M. (2004) Sexual behavior and sex-associated environmental cues activate the mesolimbic system in male rats. *Neuropsychopharmacology 29*(4): 718-730.

Bardoni, R., Takazawa, T., Tong, C. K., Choudhury, P., Scherrer, G. & MacDermott, A. B. (2013) Pre‐and postsynaptic inhibitory control in the spinal cord dorsal horn. *Annals of the New York Academy of Sciences 1279*(1): 90-96.

Bass, C. E., Grinevich, V. P., Gioia, D., Day-Brown, J. D., Bonin, K. D., Stuber, G. D., Weiner, J. L. & Budygin, E. A. (2013) Optogenetic stimulation of VTA dopamine neurons reveals that tonic but not phasic patterns of dopamine transmission reduce ethanol self-administration. *Front Behav Neurosci 7*: 173.
http://www.ncbi.nlm.nih.gov/pubmed/24324415

Becerra, L. & Borsook, D. (2008) Signal valence in the nucleus accumbens to pain onset and offset. *European journal of pain 12*(7): 866-869.

Belmaker, R. H. & Agam, G. (2008) Major depressive disorder. *New England Journal of Medicine 358*(1): 55-68.

Berridge, K. C. (2007) The debate over dopamine's role in reward: the case for incentive salience. *Psychopharmacology 191*(3): 391-431.
<Go to ISI>://000244691500002

Berridge, K. C. (2012) From prediction error to incentive salience: mesolimbic computation of reward motivation. *Eur J Neurosci 35*(7): 1124-1143.
http://www.ncbi.nlm.nih.gov/pubmed/22487042

Berridge, K. C. & Robinson, T. E. (2016) Liking, wanting, and the incentive-sensitization theory of addiction. *Am Psychol 71*(8): 670-679.
http://www.ncbi.nlm.nih.gov/pubmed/27977239

Bobzean, S. A., DeNobrega, A. K. & Perrotti, L. I. (2014) Sex differences in the neurobiology of drug addiction. *Experimental neurology 259*: 64-74.
32

Jung, C. (2016) *Psychological types*, Taylor & Francis.

Keiflin, R. & Janak, P. H. (2015) Dopamine Prediction Errors in Reward Learning and Addiction: From Theory to Neural Circuitry. *Neuron 88*(2): 247-263.
http://www.ncbi.nlm.nih.gov/pubmed/26494275

Knill, D. C. & Pouget, A. (2004) The Bayesian brain: the role of uncertainty in neural coding and computation. *TRENDS in Neurosciences 27*(12): 712-719.

Kriegeskorte, N. & Douglas, P. K. (2018) Cognitive computational neuroscience. *Nat Neurosci 21*(9): 1148-1160.

Kumar, P., Waiter, G., Ahearn, T., Milders, M., Reid, I. & Steele, J. (2008) Abnormal temporal difference reward-learning signals in major depression. *Brain 131*(8): 2084-2093.

Laasonen-Balk, T., Kuikka, J., Viinamäki, H., Husso-Saastamoinen, M., Lehtonen, J. & Tiihonen, J. (1999) Striatal dopamine transporter density in major depression. *Psychopharmacology 144*(3): 282-285.

Lagier, S., Carleton, A. & Lledo, P. M. (2004) Interplay between local GABAergic Interneurons and relay neurons generates gamma oscillations in the rat olfactory bulb. *Journal of Neuroscience 24*(18): 4382-4392.
<Go to ISI>://000221322400011

Lee, T. S. & Stella, X. Y. (2000). *An information-theoretic framework for understanding saccadic eye movements.* Paper presented at the Advances in neural information processing systems.

Lisman, J. E. & Grace, A. A. (2005) The hippocampal-VTA loop: controlling the entry of information into long-term memory. *Neuron 46*(5): 703-713.

Maslow, A. H. (1958) A Dynamic Theory of Human Motivation.

Mehta, T. R., Monegro, A., Nene, Y., Fayyaz, M. & Bollu, P. C. (2019) Neurobiology of ADHD: A Review. *Current Developmental Disorders Reports 6*(4): 235-240.

Melzack, R. & Wall, P. D. (1965) Pain mechanisms: a new theory. *Science 150*(3699): 971-979.
http://www.ncbi.nlm.nih.gov/pubmed/5320816

Messias, J. P., Paula, J. R., Grutter, A. S., Bshary, R. & Soares, M. C. (2016) Dopamine disruption increases negotiation for cooperative interactions in a fish. *Scientific reports 6*: 20817.

Mohebi, A., Pettibone, J. R., Hamid, A. A., Wong, J. T., Vinson, L. T., Patriarchi, T., Tian, L., Kennedy, R. T. & Berke, J. D. (2019) Dissociable dopamine dynamics for learning and motivation. *Nature 570*(7759): 65-70.
http://www.ncbi.nlm.nih.gov/pubmed/31118513
36